# Audio Deepfake Perceptions in College Going Populations

Gabrielle Watson, Zahra Khanjani, Vandana P. Janeja
{watson7, zkhanja, vjaneja}@umbc.edu

**Abstract**

Deepfake is content or material that is generated or manipulated using AI methods, to pass off as real. There are four different deepfake types : audio, video, image and text. In this research we focus on audio deepfakes and how people perceive it. There are several audio deepfake generation frameworks, but we chose MelGAN which is a non-autoregressive and fast audio deepfake generating framework, requiring fewer parameters. This study tries to assess audio deepfake perceptions among college students from different majors. This study also answers the question of how their background and major can affect their perception towards AI generated deepfakes. We also analyzed the results based on different aspects of: grade level, complexity of the grammar used in the audio clips, length of the audio clips, those who knew the term deepfakes and those who did not, as well as the political angle. It is interesting that the results show when an audio clip has a political connotation, it can affect what people think about whether it is real or fake, even if the content is fairly similar. This study also explores the question of how students' background and major can affect their perception towards deepfakes.

## 1 Introduction

The key differentiator between manual editing and deepfakes is that deepfakes are AI generated or AI manipulated and closely resemble real life artifacts. Deepfakes have appeared in various aspects of society including domains of politics [1, 2] , and social media [3, 4, 5, 6], and entertainment [7, 8, 9, 10, 11]. In this paper we investigate one particular category of deepfakes namely **audio deepfakes** and the perception of people towards audio deepfakes. This is particularly important as we are constantly bombarded with content on social media and we need to discern whether what we see or hear is real of fake to make informed choices. For AI generated deepfakes we used the MelGAN model Transformer TTS (TransformerTTS) [13, 14]. We selected MelGAN since it is a non-autoregressive feed forward model, which requires very few parameters as compared to other models and is relatively faster to generate the audio deepfakes[15]. Moreover, it has some easy to implement PyTorch implementations (such as [16]).

Using this model we generate audio deepfakes and evaluate perceptions of college going population towards deepfakes. We focus on the college going population since 98 % of students use social media [17], consume a large amount of web content, and are also the focus of many of the social media campaigns.

To capture perceptions towards deepfakes we conducted a survey and asked college students to listen to eight audio clips and answer if they thought an audio clips was real or fake. We also asked about their perceptions if they had heard of deepfakes and how they would perceive content in the future. We make the following **Contributions:**

- We provide a qualitative framework to evaluate perceptions towards deepfakes.

- We include audio deepfakes across different constructs of grammar, word structuring, lengths of the audio clips and political perceptions.

- Our study specifically focuses on the college going population.
- We provide brief analysis across some of the dimensions of our study including for grade level, grammar and some insights into student perceptions by majors.

Our research showed that there has only been one perception study for audio deepfakes [18], and that is for German speakers. This study used an online game simulation where 200 participants completed 8967 games with an AI algorithm that could detect audio deepfakes [18]. The study consisted of those predominately from Germany (158) so most of the data was from German speech but also included those from some other countries. The age range was from 18-85 and the ASVSpoof 2019 database was used for testing. Our study uses Transformer TTS. Also, we target an age range from 18-24 for college going, English speaking population. This was mentioned as a limitation in the prior study as they included predominately German speaking population[18]. Moreover, the study did not evaluate various aspects of grammar in audio deepfakes. We address both these aspects in our study. To the best of our knowledge our work is the first study for English speakers concerning such perceptions and assessment of discernment of audio deepfakes. Moreover, our study uniquely targets college students. The rest of the paper is organized as follows: In section 2 we present our methodology. The experimental results are discussed in section 3. Section 4 outlines the summary of our findings and future direction.

## 2    Methodology

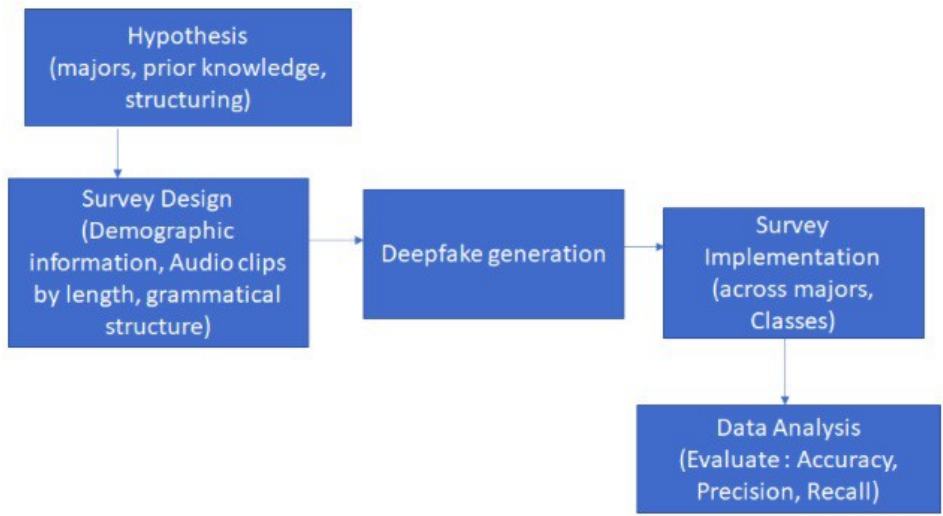

**Figure 1:** Overall Study Methodology

Our methodology includes the following steps: Based on a literature evaluation we identified hypotheses to be tested. We then designed the survey to include questions and types of clips that would help us evaluate the hypotheses. We chose a deepfake audio clip generation framework and generated the deepfake audio clips to implement the survey with the clips varying the grammar and complexity of the text to generate the clips. Required analysis and conclusions were then performed on the data obtained from the survey implementation. We next describe each step.

## 2.1 Hypothesis

We consider the following hypotheses:

- *Students who claim to have prior knowledge of deepfakes are more likely to identify deepfakes.* This is because background knowledge of a subject or knowing the term might help the user to identify the deepfake.

- *Deepfakes with longer length would be easier to identify as fake.* This is because they could hear more of the audio and if it was fake sounding such as no breaths, sighs etc. that it would be easier to detect that it was fake. Or maybe the students would think the longer the audio that it was a trick into thinking it was real because it was longer.

- *Fewer modifications in the training set or using non-complex grammar for the deepfake generator would be harder to detect as fake.* This went along with the idea that if we duplicate the sentence almost exactly that the students would not be able to tell the difference between real and fake audio.

- *Grammar and punctuation will have an impact in the ability to identify the deepfakes.* We wanted to gauge the students' reaction to complex grammar to include multiple punctuation and how the computer would say accents and umlauts. The punctuation varied greatly from the original sentence so that it could tip students off to it being fake and if the computer could not pronounce the accents and umlauts it will sound robotic and will cause them to think its fake.

- *Those who were a higher grade level would be able to discern the deepfakes better.* This is because of more exposure or have a greater analytical skills because they had taken higher level classes.

- *Students would detect the sentences affiliated to political personas as fake.* We wanted to see discernment around political leaders' names. This ties into how emotions affect our decisions and our ability to discern real from fake in this case real and fake audio clips.

We also performed analysis by computing or non-computing majors, however there was a skew in our population with a large computing major population. So, we used under sampling of computing majors to balance the skew with non-computing majors. While we discuss some preliminary findings for the comparison by majors we plan to develop additional hypothesis in a future version of this study.

## 2.2 Survey Design

We collected some basic information such as prior knowledge about deepfakes. Next, the listening portion is presented where participants listen to eight audio clips and chose completely real, unsure, or completely fake option based on their perception. After that section participants get a score of how many they got right and wrong. After this we present questions to gauge what the participants had thought and learned in the last section.

Table 1 shows the list of the clip content generated, and their true classes (real or fake). The parameters we used in creating the audio clips were word structuring, umlauts, punctuation, length of the clip variable from three to ten seconds, accents, and political candidate names used in clip.

For clips 1 and 2 tongue twisters were used because it is something that we thought could be hard for an AI to replicate matching the speed of how the tongue twister is supposed to be said. The length of the first clip was ten seconds while the length of the clip 2 was nine seconds, which was the real tongue twister so the human was able to say it faster than the model generated. Clip 3 was a copy of clip 4 generated from MelGAN. Clip 4 was the same text, but it came from the real LJ speech dataset and was chosen because of its length and its different punctation, semicolon, comma, and period. We changed the audio clip 5 to see if the users would catch on to a word change and cause them to say it was fake. Since it was longer, we assumed more people would guess this correctly than clip 6 based on length alone because maybe they could hear more of the background noise generated. The audio clip 6 was a variation of clip 3 but we were testing if the model could pronounce hard to replicate sounds. The clips 7 and 8 were not from the LJ speech dataset but one we made up to see political perceptions just by mention of political candidate names. Clip 7 was seeing if students had a personal reaction to the clip based on the name of a political candidate and also, we had a very slight variation of length. Clip 8 was one second shorter than clip 7 and mentioned another political candidate to see students' perceptions.

**Table 1:** Clip Format

| Clip Number | Fake or Real | Construct | Length in seconds |
|---|---|---|---|
| 1 | Fake | Tongue Twister | 10 |
| 2 | Real | Tongue Twister | 9 |
| 3 | Fake | No variance | 6 |
| 4 | Real | Normal sentence with commas and semicolon | 7 |
| 5 | Fake | Punctuation variation and word replacement | 8 |
| 6 | Fake | Accents and umlauts | 6 |
| 7 | Fake | political candidate shorter | 4 |
| 8 | Fake | Political candidate, shorter | 3 |

### 2.3 The Audio Deepfake Generation

**Model used:** The deepfake generator that we used was an **MelGAN** model [14, 19] run inside a Google Colab notebook called Synthesize Autoregressive MelGAN. The code was available through a Github repository [13]. The pretrained model can work with the other vocoders as well such as HiFIGAN [22], and waveRNN [23]. The MelGAN model is non-autoregressive making it fast; it is fully convolutional with significantly fewer parameters than the other models. We were able to change the input of the text to run the generator with what we wanted it to say. LJ speech (pre-trained LJSpeech models) [20] dataset was utilized, which was embedded in the notebook. After the text was synthesized the Griffin-Lim algorithm played "vocoding" role. The spectrogram was converted to a waveform . Griffin-Lim is used because it helps with consistency of a spectrogram by iterating two projections [21]. After this, the audio wave is generated by the chosen vocoder.

## 3 Experimental Results

### 3.1 Survey Data Distributions

The survey was sent out to seven computing major classes. In addition, we also recruited students from a non-computing major classes including dance vocal singing and graphic design.

There were 61 people who started the survey but only 53 made it to the audio clip questions and completed it. The distribution of majors who took it were majority IS students. The other majors who took it were Computer Engineering major, Dance, Business Technology Admin, Biology, Interdisciplinary , and Visual Arts.

The distribution of the gender, and year of study is shown in table 2. We also wanted to explore the perceptions towards deepfakes, by majors, in the broad categories of computing vs. non-computing majors. However, for this particular analysis, due to the skew in our population we extracted a smaller sample (16 respondents) to have equal numbers of computing and non-computing majors. For all other analysis we consider the full population of 53 respondents.

**Table 2:** Distribution of survey respondents

| Majors | Business Technology Admin, Biology, Computer Engineering, Dance, Information Systems, Interdisciplinary Studies, and Visual Arts |
|---|---|
| Gender Distribution | 33 males, 18 females, 2 non binary |
| Year of study | 1 Freshman, 4 sophomores, 12 juniors, 37 seniors |

## 3.2 Analysis by parameters

**Labelling of data:** We labelled the clips to study the accuracy of fake or real clip identification by students. We defined some metrics to measure the students' performance: true positive (TP) means the number of cases that are classified as fake and they are truly fake. True negative(TN) is the number of cases which are classified as real, and they are real. False negative(FN) was the number of cases that are considered as real, but they are truly fake. False Positive (FP) was the number of cases that are considered as fake, but they are real! Positive (P) is the number of all the fake clips and Negative (N) is all the real clips. Although, traditional accuracy is how accurately participants were able to guess the fakes based on the formula: $(TP + TN)/(TP + TN + FP + FN)$, however, this formula is not necessarily a suitable metric for our survey. Our survey contains a third option to be chosen which is "unsure". The traditional accuracy formula is not able to cover "unsure" cases. Therefore, we used $(TP + TN)/(P + N)$ to calculate the correctly classified portion. Also, Precision was computed as $(TP)/(TP + FP)$ and recall was computed as $(TP)/(TP + FN)$.

**Analysis by grade level**: Accuracy, Precision and Recall by grade level is shown in figure 2. We had a very low representation of Freshman and sophomores. So, we just compare seniors and juniors with each other. The seniors and juniors had the same accuracy. Ultimately, a higher grade level did not directly increase the accuracy, so our hypothesis was incorrect. However, a more comprehensive evaluation is needed across a larger population from all majors.

**Analysis by grammar:** The complex vs. non-complex clips are shown in table 2. When making the testing clips the grammar was changed substantially on question 5 and 6 to see if the respondents could tell the difference.

We categorize complex as adding multiple punctuation marks that was not there previously, adding accents or umlauts that is hard for the AI to replicate, tongue twisters, and changing words in the sentence. With this categorization there were four complex clips and four non-complex ones. The complex clips include clips 1,2,5, and 6 and non-complex is 3,4,7,and 8. Accuracy, precision and recall by complex and non-complex clips is shown in figure 3. Overall, the accuracy was higher for the complex category than the non-complex category with only a small difference in precision and a big difference in recall. So, students were able to guess more accurately the complex sentences were fake or real because of the complexity made it easier especially in the cases it was fake and there were additional punctuation and grammar like accents that made it different from the base sentence and made it sound more robotic.

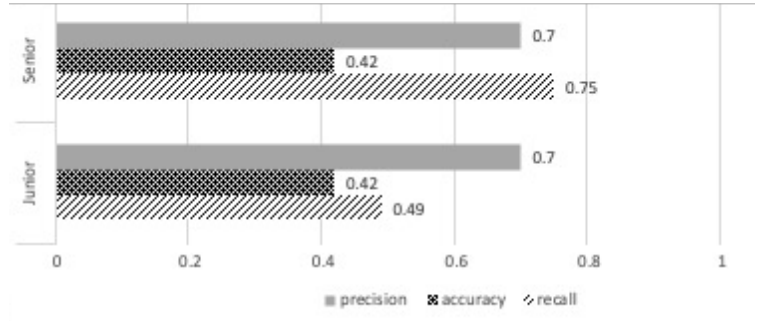

**Figure 2:** Accuracy, Precision and Recall by grade level

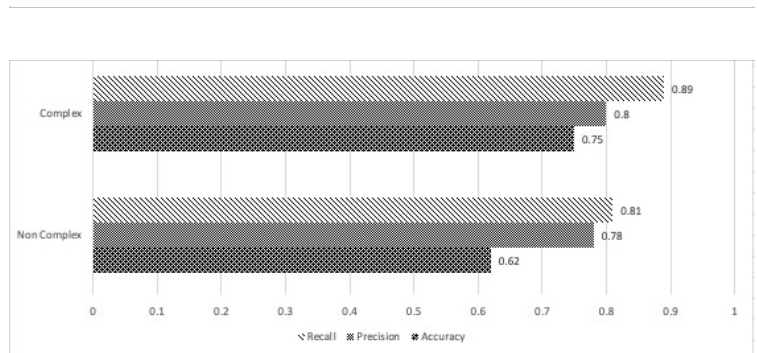

**Figure 3:** Accuracy, Precision and Recall for Complex Vs Non-Complex clips

**Analysis by length:** The way that the clips were categorized into long and short clips are based on the information which is presented in 1. All of the clips with 6 or less seconds length are considered as short, and the other ones are long. Overall the shorter clips had a significantly better accuracy, as shown in figure 4.

**Analysis by prior knowledge of deepfakes:** The data is imbalanced because those who said they knew deepfakes is much higher than those who said they did not. The places where students said they saw deepfakes was online which was mostly social media, where 34% of participants had seen deepfakes on YouTube, 19% from Reddit, 17% from Instagram, 13% from Facebook, and 17% from other. In the other category one person responded they had seen it on Twitter, and another had seen it on various websites and web players.

**Analysis by political candidate name appearing in text:** In clip 7, which was "[Candidate 1] likes burgers and fries." all respondents said it was fake. However, in clip eight "[Candidate 2] likes burgers and fries." not everyone said it was fake. For the purposes of the paper the candidate names are anonymized since we need more detailed study to understand the impact of political perceptions of the students and how they conflate with the deepfake perceptions. The clips were purposely made easy with no complicated grammar, sentence structure or hard to pronounce parts of speech so the students could more easily tell it was a fake.

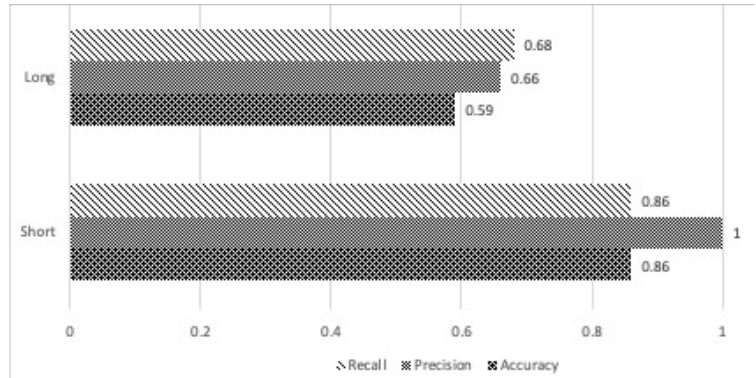

**Figure 4:** Accuracy, Precision and Recall for long vs short clips

**Analysis by Computing and Non-Computing majors:** As we can see in table 2 our dataset was imbalanced since most of the students were from computing majors, which made it hard to accurately compare one background to another. Computing majors included Information Systems (IS), Computer Engineering (COMPE), and Business Technology Administration (BTA) while non-computing majors were categorized as Biology (BIO), Dance, Individualized Study Program (INDS) and Visual Arts. We resolved our skewed dataset using under-sampling. Therefore, 7 IS students were randomly chosen, and in addition to the only COMP E and BTA students, we have a subset of 9 computing students. We compared this subset performance with the total non-computing students' performance in identifying the audio deepfakes. The overall accuracy, precision and recall were higher with the non-computing students. Overall, computing students were not any better than non-computing students similar to the study done with IT professionals not being better than people not in that career field [18]. In our case the non-computing students actually did better overall. However, there is a need to repeat this study with balanced and larger response rate. However, even with the smaller sample these findings are interesting and promising direction to explore further and understand the mechanics of how deepfakes are interpreted and discerned by different majors.

**Table 3:** Clip Format

| Major | Accuracy | Precision | Recall |
|---|---|---|---|
| Computing | 0.76 | 0.86 | 0.78 |
| Non-Computing | 0.79 | 1 | 0.81 |

## 3.3   Discussion and Future Directions

As this is one of the first perception studies for audio deepfakes focused on college students, future researchers can see how college students' perceptions change as deepfakes grow in our current times. Our hope from this study is also to bring light on this topic for educating our students about digital citizenship and maybe they will be more educated about the subject and be able to identify audio deepfakes. Future researchers could look at this work and see the framework and use that as an initial guide for their studies. The samples constructed serve as a sample of what types of speech are hard for the computer to replicate so they can also test for these kinds of speech such as tongue twisters, accents and umlauts, punctuation and a political angle to see users' perceptions. Further work could be done by asking students their political affiliation and testing solely on students' reaction to different politicized phrases. Also, one may try other high-quality audio deep fake generation frameworks such as WaveGlow even if they are not as fast and easy to implement as MelGan.

Students had better accuracy for complex sentences, shorter sentences, those who had not known what deepfakes were. School year grade does not have any impacts on the perceptions of deepfake, however, more samples across all grade levels are needed to form a significant conclusion. For the political example we did see some conflation between names of candidates and their perceptions towards deepfakes. A deeper evaluation for the policital impressions is needed.

This survey was created to make students aware of the deepfake issue, during a very uncertain time with political uncertainty and COVID. All majors are impacted by deepfakes because most college students use social media and when colleges had to switch to an online format during the pandemic, resulting in students being online even more than usual so maybe even more students could have been exposed to deepfakes than in the past in a traditional school format.

**Limitations:** A limitations of this study was that the study was conducted in two weeks during finals when many students were busy. Also, in an online environment students are already overwhelmed with emails and might have not seen this email or wanted to take this survey. Additionally, there was not an even distribution of computing and non-computing students as it was predominately computing students. Many of our students were IS majors, a total of 45 students. Also, it would have been better to have larger set of participants overall for a sample size instead of 51 so it would have been a better distribution of majors perhaps. It would have been interesting as well to see a more even distribution of the other computing majors like Computer Science, COMPE and BTA to more accurately compare computing versus non computing. We only analyzed those who lived in the U.S and could speak English however, it would be interesting to get participants who speak a different language from English from multiple countries all over the world to see what the difference would be as well in how they discern audio deepfakes. And, since our generator used the English language to pronounce speech and was trained on the LJ dataset it was limited in that sense and also could not pronounce the German umlauts and Spanish accents. We could do this by using generators trained on different languages such as Spanish so it would pronounce accents correctly. We also did not ask for political affiliation because we did not want to overcrowd the focus of the survey and make it politically inclined.  However, upon further analyses this would have been an interesting datapoint to analyze by political stance with clips seven and eight to see if that swayed which answer the participants chose.